\documentstyle[epsf]{aipproc}

\begin{document}
\begin{flushright} DOE/ER/41014-26-N97\end{flushright}
\title{Color Transparency - Color
      Coherent Effects in Nuclear
      Physics}
\author{Gerald A. Miller}
\address{Department of Physics,\\ 
         University of Washington,\\ 
         Seattle, WA 98195-1560}

\maketitle

\begin{abstract}
Efforts to observe color transparency in (e,e'p), (p,pp) and
coherent diffractive $\pi\to {\rm two\;jets}$ reactions are reviewed.
\end{abstract}

\section*{Introduction}

Color transparency is the vanishing of initial and final state interactions,
predicted by
QCD to occur in high momentum transfer quasielastic nuclear reactions.
The nuclear processes (e,e'p) and (p,pp) are examples.
The experiments need to be done with good enough resolution so that the 
basic high momentum transfer reaction (electron-proton
or proton-proton) is elastic. The energy transfer to the recoiling 
nucleus must also be small. If this is the case, the quark-gluon calculation 
of the process  is
coherent: one computes the scattering amplitude by adding all of the terms
that lead to the same  final state. The cross section is the absolute
square of that amplitude. Thus color transparency effects are  also
 color coherent effects.

Color transparency occurs\cite{mueller,brodsky} under the following three
conditions:
  \begin{itemize}
    \item  To obtain an appreciable amplitude for a high momentum transfer
          reaction on a nucleon leading to a nucleon, the colored constituents
          must be close together.\ Small objects, or 
          point-like-configurations PLCs are produced in high
          momentum transfer exclusive processes.
    \item If the quark-gluon constituents of a color singlet object
          are close together, their color electric dipole
          moment is small and the soft interactions with the medium are
          suppressed.
    \item If the small color singlet can remain small as it 
                moves through the nucleus,
    it can escape 
          without further interaction. 

          \end{itemize}
Understanding 
each of the above points represents a  serious research process; see  the 
review\cite{FMS94}. I make only 
brief remarks about each item  here.

Verifying or disproving the existence of point-like-configurations is the goal 
of color transparency
studies. Originally,  studies of  
perturbative QCD calculations of hadronic  elastic form factors
showed
that the dominant contributions came from configurations
which consisted of the fewest possible quarks (and no gluons) each at the same
impact parameter.
These calculations have been challenged. Frankfurt, Miller and
Strikman\cite{FMS92} studied the existence of point-like-configurations
 within the framework of
a wide variety of non-perturbative models. The most realistic models, 
those   which feature 
correlations between quarks, do admit the existence of 
a point-like-configuration. But the
 existence or non-existence of 
point-like-configurations can not be decided by theory-
experiments are needed.

That point-like-configurations 
have small forward scattering amplitudes is the least controversial
of the three key points. 
Such an effect occurs in theories such as QED and QCD for which there is a 
concept of neutrality. The significant  experimental 
evidence for reduced interactions comes from the existence of
scaling in low-x deep inelastic scattering, hadron-proton total cross sections
(see the review\cite{FMS94}) and in diffractive production of $\rho$ mesons
\cite{FKS97}.

The point-like-configuration
 is not a physical state; it is therefore a wave packet
which  undergoes time evolution. The point-like-configuration
initially has no size, so any 
evolution necessarily causes expansion. 
If the point-like-configuration
 expands while it is in the nucleus, it undergoes 
 typical baryonic interactions and color transparency does not occur even if
the point-like-configuration is made.
One can estimate\cite{FMS94} the laboratory expansion time
in terms of a product of a rest frame expansion time of order 1 fm
and a time dilation
factor which is expected 
 to be only about 2 for the (e,e'p) experiments
 done at SLAC and planned at TJNAF and is about 5 or 6 for the (p,pp)
experiments at BNL. Expansion effects must therefore be included to interpret
these experiments.

\section*{Current  (e,e'p) and (p,pp) data}
  
Color transparency (CT) and color
coherent effects have been recently under intense experimental and
theoretical investigation.  The (p,pp) experiment of Carroll et al.
\cite{C88} found evidence for color transparency, see Fig.~2 of
Ref.~\cite{JM93} while
the NE18 (e,e'p) experiment \cite{Makins94} did not. 
The key feature in the (p,pp) data 
 is the magnitude of the cross section, which is too large
to be explained by standard treatments fo final state interactions.
This experiment was done for $p_L$ =6, 10, and 12 GeV/c, at scattering angles
such that $Q^2= $ 4.8, 8.6 and 10.4 GeV$^2$/c$^2$. The expansion times are
roughly 3.4, 5.7 and 6.8 fm.
No such enhancement was found in the electron scattering data 
for $Q^2= $ 1, 3, 5, and 7 GeV$^2$/c$^2$.  These Q$^2$ are reasonably
large, but the outgoing
proton momenta are 1.1, 2.3, 3.5  and 4.6 GeV/c, so the expansion times
take on the small values of 
about 0.65, 1.3 2.0 and 2.6 fm.
 
The Q$^2$ of the NE18
experiment seem to be large enough to form
a small color singlet object, but the expansion times are small.
Thus this  failure to observe
significant color transparency effects is  caused by
the rapid expansion of the  point like configuration  to nearly normal
size (and  nearly normal absorption) at the relatively low  momenta
of the ejected protons \cite{F88},\cite{JM90}. In particular, models of color
transparency which reproduce the (p,2p) data and include expansion
effects predict small CT effects for the NE18 kinematics, consistent
with their findings, see the  discussion and Fig.~11 of Ref.\cite{FMS94}.
  
If this 
intepretation is correct, extending the electron scattering experiments to
a Q$^2\approx 10$ Gev$^2$/c$^2$
 such that the outgoing proton momentum is about 6 GeV/c should allow the
 observation of enhanced cross sections.

\section*{EVA}

The current (p,pp) experiment running at BNL has a new detector EVA\cite{eva}
which should allow much improved measurements. DR. S. Durant, 
explains in this
session how this detector works
and presents new preliminary data
for $p_L$=6 and 7.5 GeV/c, taken for quasielastic kinematics
such that the initial bound proton  is at rest ($\alpha=1$). They find that
the ratio of the measured to Born approximation 
cross sections $d\sigma/ d\sigma_B$ (which is unity if color transparency is
fully manifest) rises by about 30\% between those two incident momenta.

\begin{figure}
\centerline{\epsffile{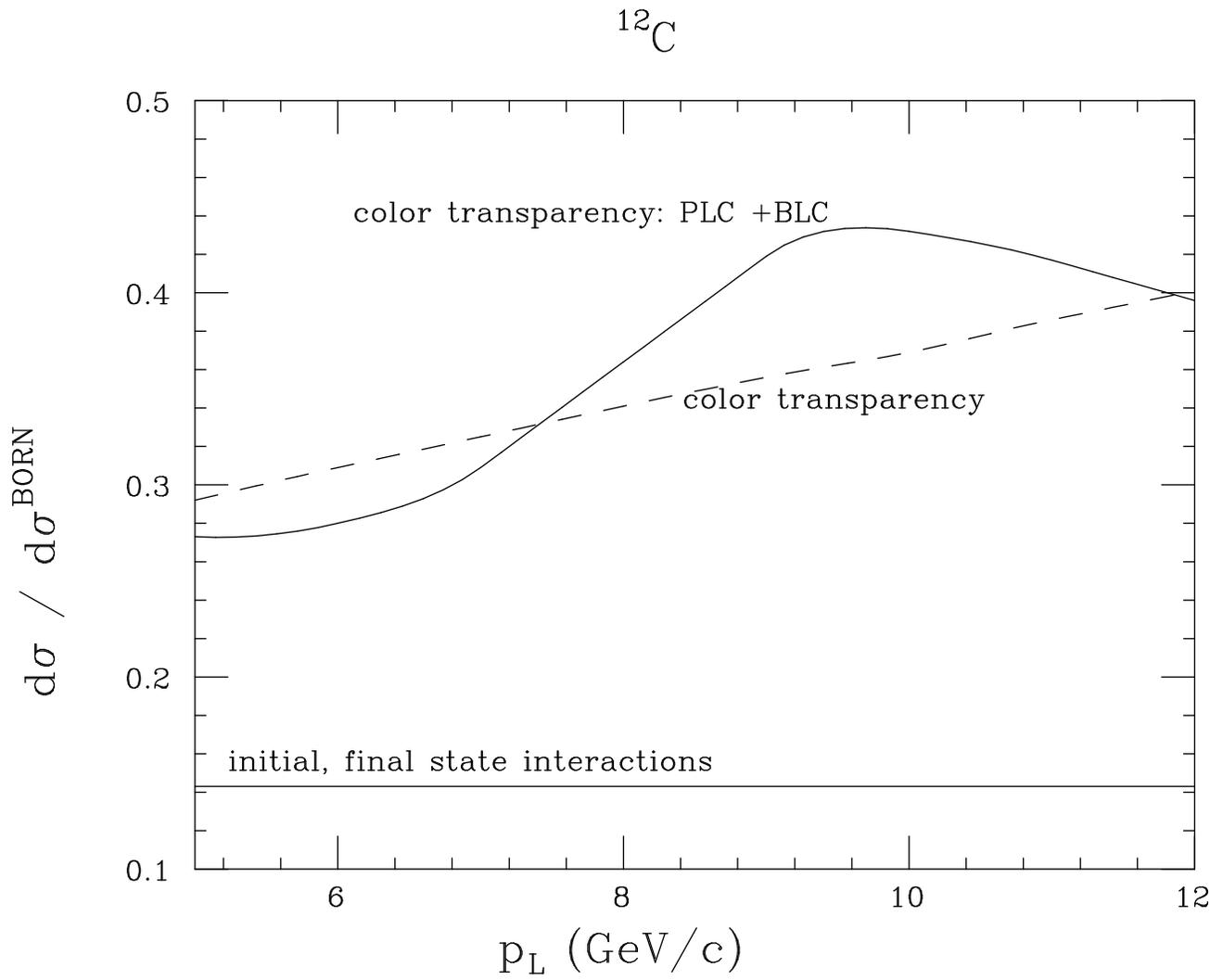}}
\vspace{-2cm}
\caption{Color transparency at quasielastic kinematics}
\label{alpha}
\end{figure}

I ran our \cite{JM93} program for this situation (before 
the meeting) and the results are displayed
in Fig.~\ref{alpha}.
 The effects of color transparency lead to an enhancement over
the usual treatment of initial and final state interactions (labelled
initial, final state interactions). The (p,pp) reaction is more complicated
than the (e,e'p) reaction because the PLC
involves all six quarks at the same point.
One expects other configurations to contribute. The simplest way to account for
such is to assume that the PLC  is complemented by another
configuration of average size, which we call the blob-like configuration.
Two models of such configurations are in the literature
\cite{brodsky2,ralston}. Here we use the version of Ref.~\cite{ralston}.
In our treatment 
such effects are not extremely important\cite{JM93} for the
kinematics of the earlier experiment\cite{JM93}, 
but seem to stand out at $\alpha=1$.
Including them leads to a more rapid rise with $p_L$ as indicated
by the preliminary data. Future data taken at higher energies will be 
very interesting. 
\section*{Victory over the expansion enemy?}

The practical problem in looking for color transparency
 effects in experiments at $Q^2$
from about one to a few GeV$^2$ is that the assumed PLC expands rapidly
while propagating through the nucleus. To observe  color transparency at
intermediate values of $Q^2$ it is necessary to suppress the effects of
wavepacket expansion.  The propagation distances are small
for the the lightest nuclear targets, but 
then, the
transparency is close to unity and the effects of CT in $(e, e'p)$
reactions tend to be 
 small.  However, if one studies a process where the produced
system can {\bf only} be produced by an interaction  in the final state,
the color coherent effects would be manifest
 as a decrease of the probability for final-state interactions
with increasing $Q^2$.   Thus, the measured
cross section would  be compared with a vanishing quantity, the
relevant ratio of cross sections would vary between unity and infinity and,
a big signal could be possible.
The first calculations\cite{double} showed that substantial CT effects 
are observable in the
(e,e'pp) reactions on $^{4,3}$He targets. Later calculations\cite{FMS95}
showed that substantial effects of color transparency are possible to observe
using the deuteron as a target in the process
$e\;{\rm d}\to e'$pn. Such experiments are planned to run at
the Jefferson Lab\cite{KG}.

Another idea involves pionic degrees of freedom. 
For some reactions involving nucleons,
 the initial and final state interactions
are expected to be dominated by exchanges of pions. 
 These interactions are also hindered in high
momentum transfer nuclear quasielastic reactions because the probability 
for a PLC to 
emit
a quark-anti-quark pair is suppressed by color neutrality. The vanishing of
pion exchange interactions has been called  \cite{FMS92}
``chiral transparency".

One example is the quasielastic production of the
$\Delta^{++}$ in electron scattering - the $(e,e^\prime\Delta^{++})$
reaction. The initial singly charged object is knocked out of the nucleus
by the virtual photon and converts to a $\Delta^{++}$ by emitting or
absorbing a charged pion. But pionic coupling to small-sized systems is
suppressed, so  this cross section for quasielastic production of
$\Delta^{++}$'s should fall faster with increasing  $Q^2$ than the
predictions of conventional theories. The first calculations\cite{FLMS97}
indicate that 
large effects are possible.
\section*{Coherent nuclear diffraction of high energy pions into two jets}

Consider 
a coherent nuclear process in 
which a high-energy pion undergoes diffractive dissociation
into a $q \bar q$ pair of high relative transverse momentum $\vec k_\perp$
but the nucleus  remains intact.   This is a very simple way to 
select a PLC in a hadronic projectile\cite{FMS93}.

  If the final $q\bar q$ pair carries of all of
pion's energy, only the
$q\bar q$ component of the light-cone wave function is needed for 
calculations. In this case, only the small sized pionic configurations pass
 through the nucleus without absorption.
Observing high $k_\perp$ jets
insures that only small $q\bar q$ separations are involved\cite{FMS93}.
The jets we consider involve high (greater than about 1 GeV/c) but
not very high values of $k_\perp$
less than about 3 or which is a huge enhancement factor for heavy nuclei.
5 GeV/c (for currently accessible energies).

The small sizes   and the large beam momentum greatly 
simplify any computations of the matrix elements.
The large beam momentum insures that 
the momentum transfer to the high momentum
$q\bar q$ system is transverse, so the $q\bar
q$-nucleon interaction is essentially independent of the longitudinal
momentum of $q\bar q$ pair. Furthermore, the effects of expansion 
should be negligible.

The dominance of configurations of small size 
is a key feature of PQCD predictions for
coherent diffraction of pions into two jets.
Thus, even in a nuclear target, color
screening implies that the coherent $q \bar
q$ system can only weakly interact. In
leading-logarithmic approximation and in the light-cone gauge
only two gluons connect the 
pion - two jet system with the nucleus.
Thus  the hadronic system propagating through the nucleus
suffers no initial-state or final-state absorption, and the
nuclear dependence of 
the $\pi +A \rightarrow $ 2 jets + A   
forward amplitude will be approximately additive in the
nucleon number $A$, so the cross section would be 
 proportional to $A^2$, a huge enhancement for heavy nuclei.
The Fermilab experiment E791\cite{ruty}, using a 500 GeV pion beam,
 will measure an integral over $t$ so that the
expected $A$ dependence is $A^{4/3}$. Preliminary indications\cite{ruty} are
that this coherent process could be measurable.

Although the vector meson (two jets) suffers no final state interactions, the
forward amplitude is not strictly additive
in nuclear number since the gluon distribution itself is shadowed. This effect
must be taken into account at very high energies; 
see the discussion in \cite{FMS94}.

\section*{Quantum invisibility}

This manuscript summarizes some of the methods that researchers are using
to observe color transparency or color coherent effects.
A  clear and convincing signal of color transparency would
represent the discovery of a new physical phenomenon.
Initial and final state interactions are the bane of a nuclear physicist's 
everyday existence- their disappearance would be a surprise indeed.

Initial and final state interactions, which usually cause the 
nucleus to act similarly to a black disk, can be said to cast a shadow behind
the nucleus. Only visible objects can cast a shadow, so that color transparency
is a quantum invisibility.

\end{document}